\documentclass[11pt]{article}
\usepackage{graphicx}
\usepackage{amsmath}
\newtheorem{definition}{Definition}

\usepackage[]{caption}
\usepackage{subcaption}
\usepackage[toc,page,titletoc]{appendix}
\usepackage[a4paper, total={6in, 9in}]{geometry}
\providecommand{\keywords}[1]
{
  \small	
  \textbf{\textit{Keywords}} #1
}
\begin{document}
\title{Efficiency of the Moscow Stock Exchange before 2022}
\author{Andrey Shternshis, Piero Mazzarisi, Stefano Marmi\\
\\
        \small Scuola Normale Superiore, Piazza dei Cavalieri 7, Pisa, Italy, 56126}
\maketitle
\begin{abstract}
This paper investigates the degree of efficiency for the Moscow Stock Exchange. A market is called efficient if prices of its assets fully reflect all available information. We show that the degree of market efficiency is significantly low for most of the months from 2012 to 2021. We calculate the degree of market efficiency by (i) filtering out regularities in financial data and (ii) computing the Shannon entropy of the filtered return time series. We have developed a simple method for estimating volatility and price staleness in empirical data, in order to filter out such regularity patterns from return time series. The resulting financial time series of stocks' returns are then clustered into different groups according to some entropy measures. In particular, we use the Kullback–Leibler distance and a novel entropy metric capturing the co-movements between pairs of stocks. By using Monte Carlo simulations, we are then able to identify the time periods of market inefficiency for a group of 18 stocks. The inefficiency of the Moscow Stock Exchange that we have detected is a signal of the possibility of devising profitable strategies, net of transaction costs. The deviation from the efficient behavior for a stock strongly depends on the industrial sector it belongs.
\end{abstract}
\section{Introduction}
When prices reflect all available information, the market is called efficient \cite{Samuelson}. One way to claim the efficiency of a market is by testing the Efficient Market Hypothesis (EMH). In the weak form, the EMH considers that the last price incorporates all the past information about market prices \cite{Fama}.
If the weak form of EMH is rejected, previous prices help to predict future prices. For traders, market efficiency means that analyzing the history of previous prices does not help to design a strategy that gives an abnormal profit. For a company issuing shares, market efficiency means that the cost of its share already reflects all information about the valuation and decisions of the company. The EMH is of great interest also in research.
Mathematical models of an asset price are usually based on the assumption that the price follows a martingale: the expected value of a future price is the current value of the price. If the EMH is rejected, there should be an estimation of the future price better than its current value. In such a case, new models should be thought.

The review of works confirming the EMH was presented by Fama in 1970 \cite{Fama} and then in 1991 \cite{Fama91}. The martingale hypothesis was also tested later. It was shown that the efficiency of a market depends on the development of the country \cite{Kim}. Also, the martingale hypothesis was confirmed on short time intervals, but may be violated on longer intervals \cite{Linton}. In addition, there is a range of strategies designed to increase an expected profit. High-frequency and algorithmic trading strategies are discussed in \cite{Mandes}. Statistical and machine learning methods for high frequency trading are reviewed in \cite{Huang}. The existence of such profitable strategies contradicts the Efficient Market Hypothesis. According to Grossman and Stiglitz \cite{Grossman}, the degree of market inefficiency determines the effort investors are willing to expend to gather and trade on information. A goal of this paper is to investigate the degree of stock market efficiency using the Shannon entropy.

Before estimating the degree of market efficiency, we need to get rid of regularities that make prices more predictable, but do not imply any profitable strategies. The methodology of filtering regularities was introduced in \cite{Calcagnile}. However, such a filtering has not usually been applied in other research works (see e.g. \cite{Molgedey,Risso08,Mensi}). In fact, deviations of price behavior from perfect randomness may be the result of some known regularity pattern, such as volatility clustering or daily seasonality, but not a signal of market inefficiency.

We process data by filtering regularities of financial time series including volatility clustering and price staleness. Price staleness is defined as a lack of price adjustments yielding 0-returns. Traders may trade less because of high transaction costs and so the price does not update. See \cite{Bandi} for more details. The price staleness produces an extra amount of 0-returns called \textit{excess 0-returns}. The other source of 0-returns in the time series is price rounding. Estimations of volatility and degree of price staleness are mutually connected: excess 0-returns appearing due to price staleness tend to underestimate volatility. At the same time, volatility estimation is needed to calculate the expected amount of 0-returns due to rounding.

One way to estimate volatility in the presence of excess 0-returns was presented in \cite{Sucarrat}. It uses expectation-maximization algorithm \cite{Dempster} to estimate returns in the places of all 0-returns and uses GARCH(1,1) model to estimate volatility \cite{Bollerslev}. The maximization of the likelihood function appearing at each step of the considered algorithm requires several parameters for numerical optimization. If the estimation of volatility is sensitive to these parameters, that are user-defined, then they may affect the entropy of returns standardized by volatility and the amount of 0-returns in the time series. In this article we suggest a modification of moving average volatility estimation that requires adjusting of the only parameter that can be defined using out-of-sample testing. The idea is to adopt a simple method for volatility estimation, so that price staleness is taken into consideration. Moreover, while estimating volatility, we filter out excess 0-returns.

The degree of market efficiency has been measured for many countries. Stock indices for 20 countries including Taiwan, Mexico, and Singapore, were considered in \cite{Risso}. The efficiency of 11 emerging markets, the US and Japan markets was measured in \cite{Cajueiro} using  the Hurst exponent and R/S statistics. The review of articles about Baltic countries was presented in \cite{Degutis}. A degree of uncertainty of Chinese \cite{Ahn}, Tunisian \cite{Mahmoud}, and Portuguese \cite{Dionisio} stock markets was also considered using entropy measures. However, the efficiency of the Russian stock market has not yet been analyzed. In this paper, we present an analysis of market efficiency based on the estimation of Shannon entropy for a group of 18 stocks of Russian companies from five industries.

Our paper introduces four original contributions in the field. First, we construct the method of filtering out heteroskedasticity and price staleness. This filtering process helps to identify a true degree of market inefficiency. Second, we calculate the degree of market inefficiency for the previous decade using monthly intervals. We conclude that the degree of market inefficiency for the Moscow Stock Exchange was greater than $80\%$. Third, we determine which pair of stocks exhibits the largest amount of inefficiency, as measured by estimating Shannon's entropy on their high frequency price time series. We show that months where the predictability of stock prices attains its maximum cluster together. We find out what behavior of stocks repeats most often for inefficient time periods. Finally, we estimate the closeness of price movements using two measures of entropy. Based on these results, we cluster together groups of stocks for which the efficient market hypothesis is rejected, thus pointing out how market inefficiency display some dependence from the financial sector they belong.

The article is organized as follows. Section~\ref{Materials and methods} describes the dataset and the methodology of filtering data regularities and calculating the Shannon entropy. Section~\ref{Results} presents the results on simulated and real data. Section~\ref{Conclusions} concludes the paper.
\section{Materials and methods}
\label{Materials and methods}
\subsection{Dataset}
\label{Dataset}
We study the Moscow Stock Exchange. We consider close prices aggregated at one-minute time scale. In particular, we select only minutes of the main trading session from 10:00 to 18:40. The time interval covers ten years from 2012 to 2021. The time period is divided into monthly time intervals. We take 18 companies, 16 of them are from five sectors: oil industry, metallurgy, banks, telecommunications, electricity. All stocks are listed in the Table~\ref{Table:dataset}\footnote{There are 2520 trading days. Assuming that there are 520 minutes in each trading day, there are 1310400 trading minutes in total. We use the Brownlees and Gallo's algorithm of an outlier detection \cite{Brownlees}. See details in Appendix~\ref{Outliers}}. All data are provided by Finam Holdings\footnote{https://www.finam.ru/}. 

\begin{table}[h]
\centering
\small
\caption{Stocks of Russian companies traded at Moscow Exchange.}
\begin{tabular}{|l|l|l|l|l|}
\hline
Ticker & Company & Sector & Size & Outliers \\ \hline
       GAZP&         Gazprom& Oil    &1307427  &50     \\ \hline
       LKOH&         Lukoil& Oil    &1287582 &192      \\ \hline
       ROSN&         Rosneft& Oil    &1270592  &130     \\ \hline
       SNGS&         Surgutneftegaz& Oil    &1211809  &11     \\ \hline
       TATN&         Tatneft& Oil    &1191390   &174    \\ \hline
       SBER&         Sberbank& Bank   &1309402  &37     \\ \hline
       VTBR&         VTB Bank& Bank   &1287330  &0     \\ \hline
       CHMF &         Severstal& Metal  &1214735   &157    \\ \hline
       NLMK&         Novolipetsk Steel& Metal  &1194324    &58   \\ \hline
       GMKN&         Nornikel& Metal  &1272769    &197   \\ \hline
       MTLR&         Mechel& Metal  &1084990  &161     \\ \hline
       MAGN&         Magnitogorsk Iron and Steel Works& Metal  &1106771 &13      \\ \hline
       MTSS&         Mobile TeleSystems& Telecommunications  &1153527 &260      \\ \hline
       RTKM&         Rostelecom& Telecommunications  &1140798 &134      \\ \hline
       
       HYDR&         RusHydro& 	Electric utility  &1252584 &0      \\ \hline
       RSTI&         Rosseti& 	Electricity  &1094244 &0      \\ \hline
       AFLT&         Aeroflot& Airline  &1083552 &123      \\ \hline
       MGNT&         Magnit& Food retailer  &1184223 &544      \\ \hline
       
\end{tabular}
\label{Table:dataset}
\captionsetup{font=scriptsize}
\caption*{For each company, we specify the ticker of stock, its sector, the size of data, and the amount of outliers removed.}
\end{table}
\subsection{Apparent inefficiencies}
To estimate a degree of market efficiency, we first should eliminate the known patterns of predictability,
such as a daily seasonality. Financial agents operating in the market tend to trade less in the middle of a day. It is reflected in prices, but again this pattern in trading volume should be filtered out to detect genuine patterns of inefficiency. Other known regularities are volatility clustering, price staleness, and microstructure noise. See Appendix~\ref{Data cleaning and whitening} for the guide of filtering out apparent inefficiencies. The contribution of this article is devising a simple method for filtering volatility clustering and price staleness. One of the methods to estimate volatility is the exponentially weighted moving average (EWMA). It is described in the next section.
\subsection{EWMA}
\label{EWMA}
We define price returns as $r_t=\ln{\left(\frac{P_t}{P_{t-1}}\right)}$, where $P_t$ is the last price available at time $t$ and $\ln()$ is the natural logarithm. In order to estimate volatility $\sigma_n$, we apply exponentially weighted moving average \cite{Hunter} of the values $\mu_1^{-1}|r_i|$, $i<n$, where $\mu_1=\sqrt{\frac{2}{\pi{}}}$.

\begin{equation}
\label{Sig1}
\bar{\sigma}_n=Sig_1(\alpha, r_{n-1},\bar{\sigma}_{n-1} )=\alpha\mu_1^{-1}|r_{n-1}|+(1-\alpha)\bar{\sigma}_{n-1}
\end{equation}

Here the fact that $E[|r_n|]=\mu_1\sigma_n$ is used and more weights are given for the more recent data. An alternative formula based on the fact that $E[r_n^2]=\sigma_n^2$ is

\begin{equation}
\label{Sig2}
\bar{\sigma}^2_n=Sig_2(\alpha, r_{n-1},\bar{\sigma}_{n-1} )=\alpha r_{n-1}^2+(1-\alpha)\bar{\sigma}_{n-1}^2.
\end{equation}

A large value of return increases the value of volatility. The current value of volatility reflects all available values of returns and changes slowly if the value of $\alpha$ is small. Usually, the smoothing parameter $\alpha$ is taken close to $0$. For instance, $\alpha=0.05$ is taken in the article \cite{Calcagnile}. The value of the parameter $\alpha$ is set to be equal $0.12$ for in-sample testing and $0.22$ for out-of-sample testing in \cite{Bollen}. Instead, Hunter \cite{Hunter} suggests to use $\alpha=0.2\pm0.1$. Using the principle of the best one-step forecasting, the smoothing parameter is set to $0.06$ for the daily data and to $0.03$ for the monthly data \cite{JPMorgan}.

We follow the approach suggested by \cite{JPMorgan} (p.~97) to find the optimal value of $\alpha$. The goal is to select the parameter $\alpha$, such that it minimizes the value of $Er_{\sigma}=\sum_i(\bar{\sigma}_i^2-r_i^2)^2$. In order to minimize $Er_{\sigma}$ as a function of the only parameter $0<\alpha<1$, we apply Brent's algorithm \cite{brent}\footnote{The method is available in Python by using the function scipy.optimize.minimize\_scalar. Alternatively, we could use the golden-section search \cite{Kiefer} that requires the boundary of search and the only parameter for the stopping criteria.}.
\subsection{Estimation of price staleness}
\label{Estimation of price staleness}
Let's define an efficient price, $P^e$, as a continuous process following a Geometric Brownian Motion.

\begin{equation*}
P^e_t=P^e_0+\int_0^t{\sigma_s P^e_s d W_s}
\end{equation*}

An observed price moves along a discrete grid. Possible price values are multiples of the tick size, $d$.

\begin{equation*}
P_t=d\cdot\left[\frac{P^e_t}{d}\right]
\end{equation*}

If the efficient price changes insignificantly, the return of the rounded price will be equal to 0. Analogically, if the return of rounded price is 0, the return of efficient price has a value close to 0. We use the Equation~\ref{staleness} to determine the probability that a 0-return is appearing due to rounding:

\begin{equation}
\label{staleness}
p_i=erf(R_{i-1})+\frac{1}{R_{i-1}\sqrt{\pi}}(\exp{(-R_{i-1}^2)}-1),
\end{equation}

where $R_i=\frac{d}{\bar{P}_i\bar{\sigma}_i\sqrt{2\Delta}}$ and $erf(x)$ is the Gaussian error function; $d$ is a tick size \footnote{We estimate the tick size using 2-steps procedure for each month. First, we find the amount of significant digits in price. Then, we determine the most frequent increment in ordered prices.}, $\Delta$ is a time step\footnote{The time step between the end and start of the main trading session is set as 1 minute. Also, we consider any time gap without trading more than 2 hours as the closure of the market. We set the time step equal to 1 minute for these gaps.}, $\bar{P}$ is a rounded price, and $\bar{\sigma}$ is an estimation of volatility \cite{Shternshis}. It is obtained by considering the probability that a price following a Geometric Brownian Motion moves less than one tick size, assuming that price increments are normally distributed.

There is another source of getting 0-returns, namely price staleness. Price staleness is a regularity that means that a fundamental (efficient) price of an asset is not updated because of economic reasons. Such a reason is, for instance, a high transaction cost, which makes transactions unprofitable for traders. See \cite{Bandi} for more details. The presence of price staleness in the data implies the existence of excess 0-returns instead of values of returns of efficient price. The excess 0-returns tend to reduce the estimation of volatility. Therefore, we need to filter out 0-returns due to price staleness and keep 0-returns due to rounding.

According to our methodology of filtering out excess 0-returns presented in \cite{Shternshis}, we save 0-returns proportionally to the probability in Eq.~\ref{staleness} and set other 0-returns as missing values. We adopt this methodology to estimate the degree of price staleness together with volatility in the next section.
\subsection{Modification of EWMA}
\label{Modification of EWMA}
In this Section, we present a modification of the EWMA that takes into consideration the effect of price staleness. Our modification of the EWMA is based on the suggestion to estimate volatility $\sigma_n$ as $\bar{\sigma}_{n-1}$ (so setting $\alpha=0$), if the value of $r_{n-1}$ is missing because of price staleness. That is, there is no new information from returns to update the value of volatility.

Initially, the expected amount of 0-returns due to rounding is $N_{save}=0$. Thus, each appearance of 0-returns does not affect the value of volatility. A 0-return is defined as a value due to rounding and is saved in the sequence if the sum of all $p_i$ (Eq.~\ref{staleness}) moves to a new integer value. Other details and the algorithm of volatility estimation can be found in Appendix \ref{appendix: Algorithm}.

We update the estimation of volatility and price staleness minute by minute. This method has the clear advantage of making possible the online inference by processing data in real time.
\subsection{The Shannon entropy}
The degree of market efficiency is assessed by computing the Shannon entropy. The entropy of a source is an average measure of the randomness of its outputs \cite{Shannon}.
\begin{definition}

Let $X = \lbrace X_1 , X_2 , .\dots\rbrace$ be a stationary random process with a finite alphabet $A$ and a measure $\mu$. An $n$-th order entropy of $X$ is
\begin{equation*}
\label{eq:entropy}
H_n(\mu)=-\sum_{x_1^n \in A^n}\mu(x_1^n)\log{\mu(x_1^n)}
\end{equation*}
with the convention $0\log{0}=0$. The process entropy (entropy rate) of $X$ is
\begin{equation*}
h(\mu)=\lim_{n\to \infty} \frac{H_n(\mu)}{n}.
\end{equation*}
\end{definition}
\subsection{Discretization}\label{section:Discretization}
The Shannon entropy is computed over a finite alphabet. To measure the Shannon entropy, we need to keep the length of blocks of symbols, $k$, sufficiently large. The predictable behavior of returns can be seen on blocks of greater length and may not be noticeable on blocks of smaller length. For this reason, we consider 3-symbols and 4-symbols discretizations using empirical quantiles.
\begin{equation*}
\begin{split}
s^{(3)}_t=
\begin{cases}
1, r_t\le\theta_1, \\
0, \theta_1<r_t\le\theta_2,\\
2,\theta_2<r_t,
\end{cases}
\end{split}
\quad
\begin{split}
s^{(4)}_t=
\begin{cases}
0, r_t\le Q_1, \\
1, Q_1<r_t\le Q_2,\\
2, Q_2<r_t\le Q_3,\\
3, Q_3<r_t,
\end{cases}
\end{split}
\end{equation*}
where $\theta_1$ and $\theta_2$ are tertiles and $Q_1$, $Q_2$, $Q_3$ are quartiles. The tertiles divide data into three equal parts. The quartiles divide data into four equal parts. $Q_2$ is also the median of the empirical distribution of returns.
For the later analysis, we will need a discretization describing the behavior of a pair of stocks.
\begin{equation}
\label{Eq:Comovement}
\begin{split}
s^{(p)}_t=
\begin{cases}
0, r^{(1)}_t\le m_1 \& r^{(
2)}_t \le m_2,\\
1, r^{(1)}_t\le m_1 \& r^{(2)}_t > m_2,\\
2, r^{(1)}_t > m_1 \& r^{(2)}_t\le m_2,\\
3, r^{(1)}_t > m_1 \& r^{(2)}_t > m_2,\\
\end{cases}
\end{split}
\end{equation}
where $r^{(1)}_t$ and $r^{(2)}_t$ are two time series of price returns and $m_1$ and $m_2$ are their medians.
\subsection{The estimation of entropy}
Let $x_1^n\in A^n$ be the sequence of length $n$ generated by an ergodic source $\mu$ from the finite alphabet $A$, where $x_i^{i+k-1}=x_i\dots x_{i+k-1}$. There are possible missing values in the sequence generated independently from $x_1^n$. We consider all blocks of length $k$ that do not contain missing values. We take

\begin{equation}
\label{value of k}
k=max(K: K<\lfloor\log(n_b(K))\rfloor,
\end{equation}

where $n_b(k)$ is the number of blocks of length $k$. The base of the logarithm is the size of the alphabet $A$ (3 or 4).

For each $a_1^k \in A^k$ empirical frequencies are defined as

$$
f(a_1^k|x_1^n)=\#\{i \in [1, n-k+1]: x_i^{i+k-1}=a_1^k \}.
$$

Empirical frequencies are the actual amount of each block from $A^k$ in the data. By considering an empirical k-block distribution as
\begin{equation}
\label{empirical prob}
\hat{\mu}_k(a_1^k|x_1^n)=\frac{f(a_1^k|x_1^n)}{n_b},
\end{equation}
an empirical $k$-entropy is defined by
\begin{equation*}
\hat{H}_k(x_1^n)=-\sum_{a_1^k}\hat{\mu}_k(a_1^k|x_1^n)\log{(\hat{\mu}_k(a_1^k|x_1^n))}=\log(n_b)-\frac{1}{n_b}\sum_{i=1}^{M}f_i\log{f_i}.
\end{equation*}

The estimation of the process entropy is

\begin{equation*}
\hat{h}_k=\frac{\hat{H}_k}{k}.
\end{equation*}
See \cite{Marton} for the proof of the consistency of this estimator and \cite{Shternshis} for the case of missing values. Since the sequence is finite, the estimation of entropy is underestimated. To remove this bias, we use the correction for the entropy estimation introduced in \cite{Grassberger, grassberger22}.

\begin{equation}
\label{Eq: Grassberger}
    \hat{H}_k^G=\log(n_b)-\frac{1}{n_b}\sum_{i=1}^{M}f_i \log{\left(\exp{G(f_i)}\right)},
\end{equation}

where the sequence $G(i)$ is defined recursively as

\begin{equation*}
    \begin{split}
        G(1)&=-\gamma-\ln(2)\\
        G(2)&=2-\gamma-\ln(2)\\
        G(2n+1)&=G(2n)\\
        G(2n+2)&=G(2n)+\frac{2}{2n+1}\text{, }n\ge1
    \end{split}
\end{equation*}

with the Euler’s constant $\gamma =0.577215\dots$.
\subsection{Detection of inefficiency}
\label{Detection of inefficiency}
We need to do three steps to determine if the time interval is efficient or not. First, we filter out apparent inefficiencies (see Appendix~\ref{Data cleaning and whitening}). Then, we estimate the entropy of the filtered return time series using Eq.~\ref{Eq: Grassberger}. Finally, we determine if the value of entropy is significantly low relative to the case of perfect randomness. We detect inefficiency in the time interval using Monte Carlo simulations. We regard a Brownian motion as absolutely unpredictable. First, we define the length of sequences as $l=n_{b}(k)+k-1$. Then, we simulate $10^4$ realizations of Brownian motions with Gaussian increments and the length $l$. For each realization, we calculate entropy using 3- and 4-symbols discretizations. Then, we find the first percentile of the obtained entropies for each discretization. These percentiles are the bounds of $99\%$ of the Confidence Interval (CI) for testing market efficiency. Finally, we define an \textit{efficiency rate} as the ratio of the entropy of the time interval and the bound of CI. If the efficiency rate is less than $1$ for at least one type of discretization, \textit{we define the time interval as inefficient}. We provide testing for inefficiency twice using different discretizations because the unique testing may not be robust. See an example in Appendix~\ref{patterns in returns that are lost in discretization}.
\subsection{Kullback–Leibler divergence}
In addition to estimating the entropy of one time series, we can also consider the difference between two time series. The Kullback–Leibler divergence \cite{Kullback} is used to measure similarity between two sequences. For two discrete probability distributions $P$ and $Q$.

$$KL(P|Q)=\sum_{i}p_i\log{\frac{p_i}{q_i}}$$

We use $p_i$ and $q_i$ as empirical probabilities obtained in Eq~\ref{empirical prob}. Since the Kullback–Leibler divergence is asymmetric, we consider the distance between two time series proposed in \cite{Benedetto}.
\begin{equation}
\label{KL}
    D(P,Q)=\frac{KL(P|Q)}{H^G(P)}+\frac{KL(Q|P)}{H^G(Q)}
\end{equation}
The greater the distance $D(P,Q)$, the more probability distributions $P$ and $Q$ differ.
\section{Results}
\label{Results}
\subsection{Simulations}
The aim of this section is to assess the accuracy of the estimation of volatility and the degree of price staleness. We will choose the method that gives the least error of the estimation for further analysis on real data. We take the following model of an observed price $\tilde{P}_t$, $t=1\dots 2N$.
\begin{equation*}
\begin{split}
P_t&=\int_0^t{\sigma_s P_s d W^1_s}\\
\tilde{P}_i&=P_i(1-B_i)+\tilde{P}_{i-1}B_i\\
pr_t&=pr_0+\int_0^t\mu_s ds+\int_0^t\nu dW^2_s
\end{split}
\quad
\begin{split}
B_i=
\begin{cases}
1\text{ with probability }pr_i\\
0\text{ with probability }1-pr_i
\end{cases}
\end{split}
\end{equation*}
where $W^1$ and $W^2$ are two independent Brownian motions
with the length of $2N$, $N=10^5$, price $P_0=100$, and $\nu=10^{-4}$. $B=1$ stands for the case when price is not updated due to price staleness (see \cite{Bandi,Kolokolov}). Prices are rounded to two digits, thus the tick size is $d=0.01$. We consider 4 choices for $pr_t$ and $\sigma_t$ listed below.
\begin{equation*}
\begin{split}
pr_t^1&=0\\
pr_t^2&=0.1+\int_0^t\nu dW^2_s\\
pr_t^3&=0.2+\int_0^t\nu dW^2_s\\
pr_t&=0.2+\int_0^t\mu^4_s ds+\int_0^t\nu dW^2_s\\
\mu^4_t&=0.8\pi/N\cos(8t\pi/N)\\
\sigma_t^1&=5\times10^{-4}\\
\sigma_t^2&\sim ARCH(1.75\times10^{-7}, 0.2, 0.1)\\
\sigma_t^3&\sim GARCH(1.25\times10^{-8}, 0.1, 0.85)\\
\sigma_t^4&\sim GARCH(1.25\times10^{-8}, 0.15, 0.8)\\
\end{split}
\end{equation*}

We divide data into two equal parts with the size $N$. The first part is a training set for finding optimal values of $\alpha$ from Equations~\ref{Sig1} and~\ref{Sig2}. The second part is a testing set for calculating errors represented in Tables~\ref{table:Results_volatility} and~\ref{table:Results_staleness} below. We compare two methods that use $Sig_1$ and $Sig_2$ for volatility estimation. We set a fixed value of alpha, $\alpha=0.05$, as a benchmark for the comparison. We also apply non-modified EMWA estimation from Section~\ref{EWMA} with selected optimal value of $\alpha$ to show the contribution of 0-filtering to the accuracy of volatility estimation. We simulate $10^3$ prices for each model.
 
Table~\ref{table:Results_volatility} represents a mean
absolute percentage error (MAPE) that is $\frac{1}{N}\sum_i|\frac{\bar{\sigma}_i-\sigma_i}{\sigma_i}|$ for six different approaches. These approaches differ in the choice of a function for volatility, the value of $\alpha$, and the presence of missing values. Table~\ref{table:Results_staleness} represents three values for each of two methods using $Sig_1$ and $Sig_2$ for the volatility estimation. The first value is the optimal value of $\alpha$. The second is $Er_N=|\frac{N_{round}N_{A}}{N_{0}N}-1|$ where $N_{A}$ is the amount of remaining non-missing returns, $N_{round}$ is the amount 0-returns that would appear due to rounding (before adding the effect of staleness in the simulated data); $N_{A}$ is the amount of non-missing returns; and $N_{0}$ is the amount of 0-returns. $Er_N$ represents the absolute error of the proportion of 0-returns that remain in the data and are defined as 0-returns due to rounding. The third value is the proportion of data set as missing values, that is $1-\frac{N_{A}}{N}$.

It can be seen from Table~\ref{table:Results_volatility} that the method that more often gives the lowest value of MAPE is with fixed $\alpha=0.05$ and $Sig_1$ used for volatility estimation. Moreover, for almost all cases, 0-filtering makes the volatility estimate more accurate. The error of the amount of 0-returns due to rounding is smaller for the function $Sig_1$ than for the function $Sig_2$ for all 16 cases. After the comparison of the two functions of volatility estimation, we choose $Sig_1$ that uses absolute values of returns. For the rest of the paper, we fix the value of $\alpha$ as $0.05$ for the simplicity of further analysis.
\subsection{Moscow Stock Exchange}
We define a degree of inefficiency as the fraction of months which are defined as inefficient according to Section~\ref{Detection of inefficiency}. The degree of inefficiency for the chosen group of stocks traded at Moscow Exchange is 0.823.\footnote{In our previous work, \cite{Shternshis} we found that the degree of inefficiency for the U.S. ETF market is about 0.11 for monthly time intervals and the 3-symbols discretization only.} The degree of inefficiency for each stock and discretization is presented in Table~\ref{Real data results}. We notice that the 4-symbols discretization contributes to the larger amount of inefficient months than the 3-symbols discretization. That is, the 4-symbol discretization appears to have a more predictable structure than the 3-symbols discretization.

\begin{figure}[b]
\centering
\includegraphics[width=10cm]{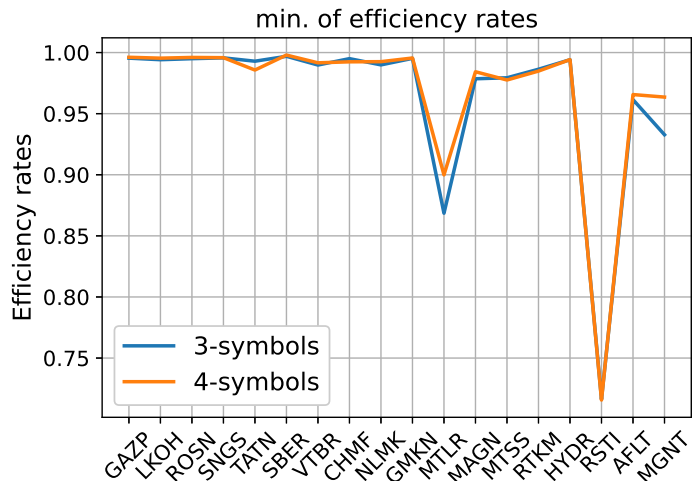}
\captionsetup{font=scriptsize}
\caption{Minimum of efficiency rates for 18 stocks using 3- and 4-symbols discretizations.}
\label{fig:efficiency_rates_1}
\end{figure}

Figure~\ref{fig:efficiency_rates_1} shows the minimum value of efficiency rate among all months for each stock.

There are two most notable deviations from 1 for stocks MLTR (Mechel, mining and metals company) and RSTI (Rosseti, power company). We investigate them in the next section. For the other 16 stocks, the minimum value of efficiency rate is attained for the stock AFLT and is equal to 0.933 (0.964) for 3 (4) symbols.

\begin{table}[hb]
\small
\caption{Results on volatility estimation.}
\centering
\begin{tabular}{|p{1cm}|p{2cm}|p{2cm}|p{2cm}|p{2cm}|p{2cm}|p{2cm}|}
\hline
model &MAPE, method $v_1$ & MAPE, $v_2$ & MAPE with $\alpha=0.05 $, $v_1$ & MAPE with $\alpha=0.05$, $v_2$ &MAPE w/o 0-filtering, $v_1$ &    MAPE w/o 0-filtering, $v_2$                                                \\ \hline
$\sigma_1,pr_1$ & $\underset{(0.0007, 0.0507)}{0.0193}$ &$\underset{(0.0014, 0.0406)}{\textbf{0.017}}$ & $\underset{(0.0955,0.0995)}{0.0975}$&$\underset{(0.0878, 0.0915)}{0.0897}$ & $\underset{(0.0007,0.0507)}{0.0193}$ &$\underset{(0.0014, 0.0406)}{\textbf{0.017}}$ \\ \hline
$\sigma_1,pr_2$ & $\underset{( 0.0245 ,  0.1017 )}{ \textbf{0.0607} }$ & $\underset{( 0.0293 ,  0.1057 )}{ 0.0629 }$ & $\underset{( 0.093 ,  0.0972 )}{ 0.095 }$ & $\underset{( 0.0893 ,  0.0936 )}{ 0.0914 }$ & $\underset{( 0.0459 ,  0.131 )}{ 0.0862 }$ & $\underset{( 0.0294 ,  0.1154 )}{ 0.0674 }$ \\ \hline
$\sigma_1,pr_3$ & $\underset{( 0.0333 ,  0.1278 )}{ \textbf{0.0737} }$ & $\underset{( 0.033 ,  0.1338 )}{ 0.0756 }$ & $\underset{( 0.0928 ,  0.0971 )}{ 0.0948 }$ & $\underset{( 0.0894 ,  0.094 )}{ 0.0915 }$ & $\underset{( 0.0917 ,  0.1863 )}{ 0.138 }$ & $\underset{( 0.0368 ,  0.1592 )}{ 0.0888 }$ \\ \hline
$\sigma_1,pr_4$ & $\underset{( 0.0323 ,  0.1213 )}{ \textbf{0.0716} }$ & $\underset{( 0.0354 ,  0.1268 )}{ 0.0739 }$ & $\underset{( 0.0926 ,  0.0973 )}{ 0.0949 }$ & $\underset{( 0.089 ,  0.0937 )}{ 0.0913 }$ & $\underset{( 0.1022 ,  0.1875 )}{ 0.1404 }$ & $\underset{( 0.0405 ,  0.1516 )}{ 0.0873 }$ \\ \hline
$\sigma_2,pr_1$ & $\underset{(0.1082, 0.121)}{0.1121}$ &$\underset{(0.1146, 0.1244)}{0.1183}$ & $\underset{(0.1438,0.1481)}{0.1459}$&$\underset{(0.1422, 0.147)}{0.1446}$ & $\underset{(0.108,0.1207)}{\textbf{0.1118}}$ &$\underset{(0.1144, 0.1243)}{0.1179}$ \\ \hline
$\sigma_2,pr_2$ & $\underset{( 0.1163 ,  0.1715 )}{ 0.1359 }$ & $\underset{( 0.1237 ,  0.1765 )}{ 0.1411 }$ & $\underset{( 0.1439 ,  0.1487 )}{ 0.1462 }$ & $\underset{( 0.1457 ,  0.1526 )}{ 0.1489 }$ & $\underset{( 0.1043 ,  0.1819 )}{ \textbf{0.1341} }$ & $\underset{( 0.1193 ,  0.1832 )}{ 0.1407 }$  \\ \hline
$\sigma_2,pr_3$ & $\underset{( 0.1198 ,  0.1958 )}{ \textbf{0.146} }$ & $\underset{( 0.1266 ,  0.1981 )}{ 0.1519 }$ & $\underset{( 0.1449 ,  0.1499 )}{ 0.1473 }$ & $\underset{( 0.1464 ,  0.1534 )}{ 0.1496 }$ & $\underset{( 0.1196 ,  0.2271 )}{ 0.1649 }$ & $\underset{( 0.123 ,  0.222 )}{ 0.1589 }$ \\ \hline
$\sigma_2,pr_4$ & $\underset{( 0.1205 ,  0.1912 )}{ \textbf{0.146} }$ & $\underset{( 0.1256 ,  0.1986 )}{ 0.15 }$ & $\underset{( 0.1447 ,  0.1498 )}{ 0.1472 }$ & $\underset{( 0.1463 ,  0.1532 )}{ 0.1494 }$ & $\underset{( 0.1274 ,  0.2261 )}{ 0.1696 }$ & $\underset{( 0.1223 ,  0.2239 )}{ 0.1571 }$ \\ \hline
$\sigma_3,pr_1$ & $\underset{(0.1446, 0.1513)}{0.1479}$ &$\underset{(0.1442, 0.1505)}{\textbf{0.1473}}$ & $\underset{(0.1467,0.1522)}{0.1495}$&$\underset{(0.1442, 0.1502)}{\textbf{0.1473}}$ & $\underset{(0.1446,0.1513)}{0.1479}$ &$\underset{(0.1441, 0.1503)}{\textbf{0.1472}}$ \\ \hline
$\sigma_3,pr_2$ & $\underset{( 0.1485 ,  0.1891 )}{ 0.1592 }$ & $\underset{( 0.1508 ,  0.1857 )}{ 0.1613 }$ & $\underset{( 0.149 ,  0.1574 )}{ \textbf{0.1529} }$ & $\underset{( 0.1497 ,  0.1598 )}{ 0.1546 }$ & $\underset{( 0.144 ,  0.2033 )}{ 0.1622 }$ & $\underset{( 0.1491 ,  0.1978 )}{ 0.1628 }$\\ \hline
$\sigma_3,pr_3$ & $\underset{( 0.1528 ,  0.2048 )}{ 0.1681 }$ & $\underset{( 0.1556 ,  0.2178 )}{ 0.171 }$ & $\underset{( 0.1525 ,  0.1616 )}{ \textbf{0.1567} }$ & $\underset{( 0.1536 ,  0.1639 )}{ 0.1584 }$ & $\underset{( 0.154 ,  0.2477 )}{ 0.1904 }$ & $\underset{( 0.1546 ,  0.2464 )}{ 0.1815 }$ \\ \hline
$\sigma_3,pr_4$ & $\underset{( 0.1527 ,  0.1997 )}{ 0.1668 }$ & $\underset{( 0.1555 ,  0.2178 )}{ 0.1701 }$ & $\underset{( 0.1525 ,  0.1613 )}{ \textbf{0.1568} }$ & $\underset{( 0.1537 ,  0.1633 )}{ 0.1583 }$ & $\underset{( 0.1591 ,  0.246 )}{ 0.192 }$ & $\underset{( 0.1556 ,  0.2455 )}{ 0.181 }$ \\ \hline
$\sigma_4,pr_1$ &  $\underset{(0.1856, 0.1952)}{0.1897}$ &$\underset{(0.1838, 0.1911)}{\textbf{0.1873}}$ & $\underset{(0.1844,0.1918)}{0.1881}$&$\underset{(0.1879, 0.1968)}{0.1924}$ & $\underset{(0.1856,0.1952)}{0.1897}$ &$\underset{(0.1837, 0.1911)}{0.1873}$ \\ \hline
$\sigma_4,pr_2$ & $\underset{( 0.1906 ,  0.2454 )}{ 0.2035 }$ & $\underset{( 0.1921 ,  0.2474 )}{ 0.2057 }$ & $\underset{( 0.1891 ,  0.2022 )}{ \textbf{0.1954} }$ & $\underset{( 0.1961 ,  0.2119 )}{ 0.2037 }$ & $\underset{( 0.1836 ,  0.2617 )}{ 0.2049 }$ & $\underset{( 0.1902 ,  0.2642 )}{ 0.2079 }$ \\ \hline
$\sigma_4,pr_3$& $\underset{( 0.1965 ,  0.2623 )}{ 0.2146 }$ & $\underset{( 0.1996 ,  0.2757 )}{ 0.2166 }$ & $\underset{( 0.1951 ,  0.2077 )}{ \textbf{0.2015} }$ & $\underset{( 0.2026 ,  0.2177 )}{ 0.2101 }$ & $\underset{( 0.1912 ,  0.307 )}{ 0.2318 }$ & $\underset{( 0.1988 ,  0.3082 )}{ 0.2294 }$ \\ \hline
$\sigma_4,pr_4$ & $\underset{( 0.1967 ,  0.2591 )}{ 0.214 }$ & $\underset{( 0.1986 ,  0.2689 )}{ 0.2155 }$ & $\underset{( 0.1951 ,  0.2088 )}{ \textbf{0.2013} }$ & $\underset{( 0.2023 ,  0.2185 )}{ 0.2097 }$ & $\underset{( 0.1976 ,  0.3064 )}{ 0.2338 }$ & $\underset{( 0.1988 ,  0.306 )}{ 0.2286 }$ \\ \hline
\end{tabular}
\captionsetup{font=scriptsize}
\caption*{The first column indicated a model. Columns 2 and 3 represent results for two methods described in Section~\ref{Modification of EWMA}. Columns 4 and 5 are for the same methods but with the fixed value of $\alpha$. Columns 6 and 7 shows the error of the standard EMWA approach with the optimal selected value of $\alpha$. $95\%$ CI is presented below each averaged statistic. $v_1$ stands for using $Sig_1$; $v_2$ stands for using $Sig_2$.}
\label{table:Results_volatility}
\end{table}

\begin{table}[ht]
\small
\caption{Results on filtering out 0-returns.}
\centering
\begin{tabular}{|p{1cm}|p{2cm}|p{2cm}|p{2cm}|p{2cm}|p{2cm}|p{2cm}|}
\hline
model &$\alpha$ for $v_1$ & $\alpha$ for $v_2$ & $Er_N$, $v_1$ & $Er_N$, $v_2$ &Fraction of data deleted, $v_1$ &  Fraction of data deleted, $v_2$                                                \\ \hline
$\sigma_1,pr_1$ & $\underset{(0.0, 0.0137)}{0.0027}$ &$\underset{(0.0, 0.0103)}{0.0022}$ & $\underset{(0.0,0.0)}{0.0006}$&$\underset{(0.0, 0.0259)}{0.0015}$ & $\underset{(0.0,0.0)}{0.0001}$ &$\underset{(0.0, 0.0052)}{0.0003}$ \\ \hline
$\sigma_1,pr_2$ & $\underset{( 0.0033 ,  0.0569 )}{ 0.0228 }$ & $\underset{( 0.0044 ,  0.067 )}{ 0.0259 }$ & $\underset{( 0.0004 ,  0.026 )}{ 0.0094 }$ & $\underset{( 0.0005 ,  0.0295 )}{ 0.011 }$ & $\underset{( 0.0814 ,  0.3244 )}{ 0.2005 }$ & $\underset{( 0.0818 ,  0.3247 )}{ 0.2008 }$ \\ \hline
$\sigma_1,pr_3$ & $\underset{( 0.006 ,  0.0902 )}{ 0.0335 }$ & $\underset{( 0.0058 ,  0.1063 )}{ 0.0379 }$ & $\underset{( 0.0005 ,  0.0288 )}{ 0.0106 }$ & $\underset{( 0.0005 ,  0.0336 )}{ 0.0121 }$ & $\underset{( 0.2474 ,  0.481 )}{ 0.3661 }$ & $\underset{( 0.246 ,  0.4797 )}{ 0.3659 }$ \\ \hline
$\sigma_1,pr_4$& $\underset{( 0.0056 ,  0.0824 )}{ 0.0314 }$ & $\underset{( 0.007 ,  0.0966 )}{ 0.036 }$ & $\underset{( 0.0004 ,  0.0283 )}{ 0.0104 }$ & $\underset{( 0.0007 ,  0.0355 )}{ 0.0122 }$ & $\underset{( 0.2521 ,  0.4717 )}{ 0.3628 }$ & $\underset{( 0.2515 ,  0.4713 )}{ 0.3626 }$ \\ \hline
$\sigma_2,pr_1$ & $\underset{(0.0, 0.0161)}{0.0039}$ &$\underset{(0.0006, 0.0146)}{0.0037}$ & $\underset{(0.0,0.0438)}{0.0149}$&$\underset{(0.0, 0.0586)}{0.035}$ & $\underset{(0.0,0.0092)}{0.0029}$ &$\underset{(0.0, 0.0134)}{0.0067}$ \\ \hline
$\sigma_2,pr_2$ & $\underset{( 0.0059 ,  0.0903 )}{ 0.035 }$ & $\underset{( 0.0054 ,  0.1021 )}{ 0.0367 }$ & $\underset{( 0.0012 ,  0.0448 )}{ 0.0209 }$ & $\underset{( 0.0039 ,  0.0601 )}{ 0.0319 }$ & $\underset{( 0.0856 ,  0.3249 )}{ 0.2016 }$ & $\underset{( 0.0878 ,  0.3263 )}{ 0.2032 }$  \\ \hline
$\sigma_2,pr_3$ & $\underset{( 0.0079 ,  0.1326 )}{ 0.0489 }$ & $\underset{( 0.0091 ,  0.1476 )}{ 0.0556 }$ & $\underset{( 0.001 ,  0.0473 )}{ 0.0217 }$ & $\underset{( 0.0022 ,  0.0603 )}{ 0.0275 }$ & $\underset{( 0.25 ,  0.4835 )}{ 0.3706 }$ & $\underset{( 0.2518 ,  0.4836 )}{ 0.371 }$
 \\ \hline
$\sigma_2,pr_4$ & $\underset{( 0.0093 ,  0.1268 )}{ 0.049 }$ & $\underset{( 0.0082 ,  0.1484 )}{ 0.0525 }$ & $\underset{( 0.0012 ,  0.0443 )}{ 0.0206 }$ & $\underset{( 0.0013 ,  0.0571 )}{ 0.0274 }$ & $\underset{( 0.2495 ,  0.4695 )}{ 0.3645 }$ & $\underset{( 0.2491 ,  0.4692 )}{ 0.3651 }$ \\ \hline
$\sigma_3,pr_1$ & $\underset{(0.0349, 0.0527)}{0.0424}$ &$\underset{(0.0392, 0.0603)}{0.048}$ & $\underset{(0.0,0.0337)}{0.0034}$&$\underset{(0.0, 0.0402)}{0.0089}$ & $\underset{(0.0,0.0067)}{0.0007}$ &$\underset{(0.0, 0.0085)}{0.0018}$ \\ \hline
$\sigma_3,pr_2$ & $\underset{( 0.0267 ,  0.1456 )}{ 0.0672 }$ & $\underset{( 0.0249 ,  0.1592 )}{ 0.0767 }$ & $\underset{( 0.0006 ,  0.0404 )}{ 0.0155 }$ & $\underset{( 0.0009 ,  0.0551 )}{ 0.02 }$ & $\underset{( 0.0826 ,  0.3248 )}{ 0.1995 }$ & $\underset{( 0.0826 ,  0.3251 )}{ 0.1997 }$ \\ \hline
$\sigma_3,pr_3$ & $\underset{( 0.0264 ,  0.1734 )}{ 0.0825 }$ & $\underset{( 0.0301 ,  0.2371 )}{ 0.0985 }$ & $\underset{( 0.0011 ,  0.0477 )}{ 0.018 }$ & $\underset{( 0.0008 ,  0.0702 )}{ 0.0243 }$ & $\underset{( 0.2444 ,  0.4746 )}{ 0.3678 }$ & $\underset{( 0.2421 ,  0.4751 )}{ 0.3671 }$ \\ \hline
$\sigma_3,pr_4$ & $\underset{( 0.0266 ,  0.163 )}{ 0.0788 }$ & $\underset{( 0.0325 ,  0.2362 )}{ 0.0969 }$ & $\underset{( 0.001 ,  0.0463 )}{ 0.0178 }$ & $\underset{( 0.0007 ,  0.067 )}{ 0.0222 }$ & $\underset{( 0.2466 ,  0.476 )}{ 0.3623 }$ & $\underset{( 0.2486 ,  0.4739 )}{ 0.3615 }$ \\ \hline
$\sigma_4,pr_1$& $\underset{(0.0696, 0.1037)}{0.0819}$ &$\underset{(0.0757, 0.119)}{0.0904}$ & $\underset{(0.0,0.0248)}{0.0013}$&$\underset{(0.0, 0.0329)}{0.0047}$ & $\underset{(0.0,0.0052)}{0.0003}$ &$\underset{(0.0, 0.0075)}{0.0011}$\\ \hline
$\sigma_4,pr_2$ & $\underset{( 0.0534 ,  0.2359 )}{ 0.1132 }$ & $\underset{( 0.0576 ,  0.2925 )}{ 0.1339 }$ & $\underset{( 0.0007 ,  0.0564 )}{ 0.0185 }$ & $\underset{( 0.0008 ,  0.087 )}{ 0.0265 }$ & $\underset{( 0.0765 ,  0.3287 )}{ 0.1993 }$ & $\underset{( 0.077 ,  0.3257 )}{ 0.1982 }$ \\ \hline
$\sigma_4,pr_3$ & $\underset{( 0.0557 ,  0.2678 )}{ 0.1338 }$ & $\underset{( 0.0597 ,  0.3734 )}{ 0.1596 }$ & $\underset{( 0.0009 ,  0.0621 )}{ 0.0214 }$ & $\underset{( 0.001 ,  0.1119 )}{ 0.0321 }$ & $\underset{( 0.2419 ,  0.4823 )}{ 0.3687 }$ & $\underset{( 0.2378 ,  0.4817 )}{ 0.3669 }$ \\ \hline
$\sigma_4,pr_4$ & $\underset{( 0.0571 ,  0.263 )}{ 0.1317 }$ & $\underset{( 0.0613 ,  0.3541 )}{ 0.1556 }$ & $\underset{( 0.0008 ,  0.0667 )}{ 0.0211 }$ & $\underset{( 0.0011 ,  0.108 )}{ 0.0315 }$ & $\underset{( 0.2599 ,  0.4823 )}{ 0.3641 }$ & $\underset{( 0.2564 ,  0.4822 )}{ 0.3625 }$\\ \hline
\end{tabular}
\captionsetup{font=scriptsize}
\caption*{Values of $\alpha$, errors of the number of 0-returns due to rounding, and fraction of data set as missing values. The first column indicated a model. $95\%$ CI is presented below each averaged statistic. $v_1$ stands for using $Sig_1$; $v_2$ stands for using $Sig_2$.}
\label{table:Results_staleness}
\end{table}

\clearpage

\subsubsection{Analysis of MLTR and RSTI}
\begin{figure}[htb]
\centering
\includegraphics[width=9 cm]{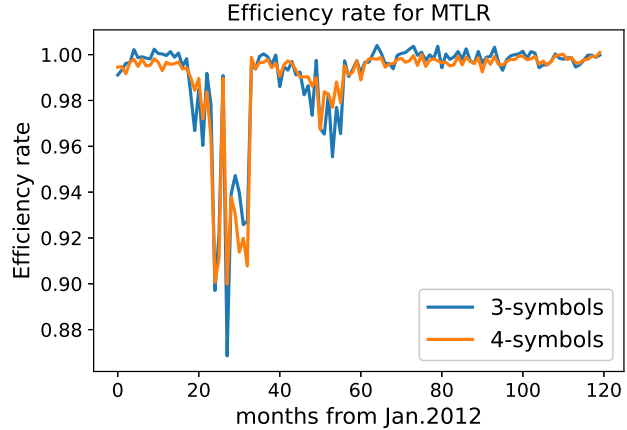}
\captionsetup{font=scriptsize}
\caption{Efficiency rate for the MLTR stock using 3- and 4-symbols discretizations.\label{fig:MLTR_ineffrate_1}}
\end{figure}

\begin{figure}[b]
\centering
\includegraphics[width=9 cm]{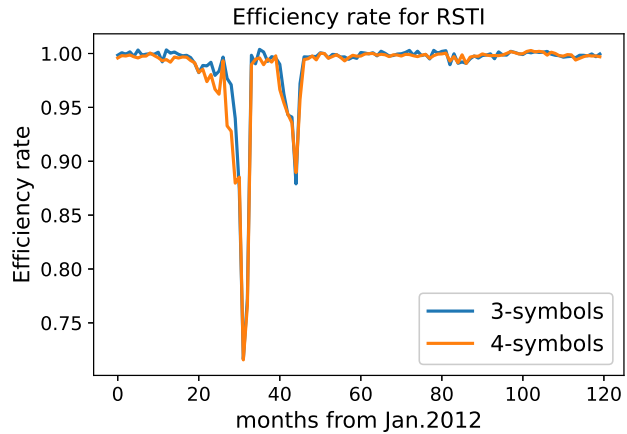}
\captionsetup{font=scriptsize}
\caption{Efficiency rate for the RSTI stock using 3- and 4-symbols discretizations.\label{fig:RSTI_ineffrate_1}}
\end{figure}

We plot the values of efficiency rates for monthly intervals for the MLTR and RSTI stocks. See Fig.~\ref{fig:MLTR_ineffrate_1} and Fig.~\ref{fig:RSTI_ineffrate_1}.

Both types of discretization show coherent results. For MLTR, there are two notable decreases in the efficiency rates at the beginning of 2014 and in the middle of 2016. For both types of discretizations, eight months with the lowest efficiency rate (in the ascending order of time) are Jan-Feb and May-Oct of 2014. For each month we write down the most frequent block of symbols in Table~\ref{table:frequentblocksMLTR}. Note that block $1111$ for the 4-symbols discretization appears as the most frequent for 6 months out of 8 for MLTR. The block denotes a slight decrease in the price for 4 minutes in a row. The meaning of the last two columns is discussed later.

For RSTI, there are two sharp decreases in 2014 and 2015. There are 11 months that have the lowest efficiency rates in common for both discretizations. These months are Apr-Sep of 2014 and Jun-Oct of 2015. Note that these inefficient months cluster together and are not distributed uniformly among the entire time period of 10 years. This is the signal of a market condition that affects the inefficiency of the stocks for more than one month.

We construct a simple trading strategy on discretized returns to test the predictability of future returns. We consider blocks of length 4 obtained by the 4-symbols discretization. For each month, we divide blocks into two equal parts. The discretization is made using only the first part of a month. We consider the sequences of the first 3 symbols of each block. If the empirical probability of getting 0 or 1 after the sequence of 3 symbols in the first part of the month, this sequence is from group D (decreasing). If the empirical probability of getting 2 or 3 after the sequence of 3 symbols, this sequence is from group I (increasing). Then, for the second part of the month, we determine a success if symbols 0 or 1 follow a sequence from group D or if symbols 2 or 3 follow a sequence from group I. Then, we calculate the fraction of successes. Thus, it is the probability of making a profit: sell after group D or buy after group I. In the case of market efficiency, this probability is equal to 0.5. For example, we expect that after $111$ the next symbol would be $1$ according to Table~\ref{table:frequentblocksMLTR}. That is, after this block, a trader can sell a stock. The fourth column of the Table~\ref{table:frequentblocksMLTR} shows the results for filtered return time series. The fifth column stands for the original return time series.

For all cases the probability is greater than 0.5. Obviously, the probabilities for the original return time series are greater than those for the filtered return time series. The reason is that predictability for the original return time series follows from the sources of apparent inefficiencies.

The same analysis is done for the RSTI stock. Eleven months with the lowest efficiency rates are presented in Table~\ref{table:frequentblocksRSTI}. For the RSTI stock, the simple trading strategy gives the fraction of successes (of predicting increases and decreases of the price) greater than $0.5$ for all 11 months. The frequent behavior of the price of RSTI during the chosen months is a slight increase in price for several minutes in a row denoted by symbol 2.
\begin{table}[t]
\centering
\caption{The degree of inefficiency for each stock.}
\begin{tabular}{|l|l|l|l|l|l|}
\hline
Ticker & Degree of inefficiency & For 3 symbols only & For 4 symbols only \\ \hline
       GAZP&         0.725& 0.392    &0.675     \\ \hline
       LKOH&         0.65& 0.342    &0.542      \\ \hline
       ROSN&         0.742& 0.392    &0.708      \\ \hline
       SNGS&         0.725& 0.4    &0.625      \\ \hline
       TATN&         0.617& 0.392    &0.525     \\ \hline
       SBER&         0.725& 0.433   &0.658      \\ \hline
       VTBR&         0.842& 0.592   &0.792       \\ \hline
       CHMF &         0.858& 0.55  &0.692     \\ \hline
       NLMK&         0.8& 0.467  &0.692     \\ \hline
       GMKN&         0.733&     0.475&0.608     \\ \hline
       MTLR&         0.992& 0.783  &0.975       \\ \hline
       MAGN&        0.833  &0.65 &0.758      \\ \hline
       MTSS&         0.967&     0.7&0.942     \\ \hline
       RTKM&         0.942&     0.683&0.908     \\ \hline
       HYDR&         0.892&     0.75&0.8     \\ \hline
       RSTI&         0.917&     0.742&0.875     \\ \hline
       AFLT&         0.983&     0.775&0.95     \\ \hline
       MGNT&         0.842& 0.667    &0.742     \\ \hline
\end{tabular}
\label{Real data results}
\captionsetup{font=scriptsize}
\caption*{Fraction of inefficient months using 3-symbols and 4-symbols discretization.}
\end{table} 
\begin{table}[ht]
\centering
\caption{The most frequent blocks appearing for the Stock MLTR and probabilities of success.}
\begin{tabular}{|p{2cm}|p{2cm}|p{2cm}|p{2cm}|p{2cm}|}
\hline
Months of 2014 & The most frequent block, 3-s & The most frequent block, 4-s & prob. of success, filtered & prob. of success, original\\ \hline
Jan.           & 00000                        & 1111 &  0.64   &0.75                   \\ \hline
Feb.           & 00000                        & 2222  &  0.64   &0.74                \\ \hline
May            & 00000                        & 1111   & 0.61   &0.73            \\ \hline
June           & 22222                        & 1111    & 0.60    &0.73       \\ \hline
July           & 11111                        & 1111     & 0.62  &0.74     \\ \hline
Aug.           & 00000                        & 1111      & 0.61 &0.76  \\ \hline
Sep.           & 00000                        & 1111       &0.63    &0.74              \\ \hline
Oct.           & 120120                       & 0303        & 0.55     &0.6         \\ \hline
\end{tabular}
\captionsetup{font=scriptsize,width=0.8\linewidth}
\caption*{The first column represents months with the lowest efficiency rates. Columns 2 and 3 are the most frequent blocks in 3- and 4-symbols discretization. Columns 4 and 5 are the probability of the success of the simple trading strategy for filtered and original price returns.}
\label{table:frequentblocksMLTR}
\end{table}
\begin{table}[ht]
\center
\caption{The most frequent blocks appearing for the Stock RSTI and probabilities of success.}
\begin{tabular}{|p{2cm}|p{2cm}|p{2cm}|p{2cm}|p{2cm}|}
\hline
Months & The most frequent block, 3-s & The most frequent block, 4-s & prob. of success, filtered & prob. of success, original\\ \hline
Apr.2014          &212121                         &0111  &  0.63   &0.77                   \\ \hline
May.2014          &00000                         &1111  &  0.61   &0.73                   \\ \hline
June.2014          &00000                         &1111  &  0.6   &0.73                   \\ \hline
July.2014          &00000                         &2222  &  0.62   &0.74                   \\ \hline
Aug.2014          &00000                         &2222  &  0.61   &0.76                   \\ \hline
Sep.2014          &000000                         &22222  &  0.63   &0.74                   \\ \hline
June.2015          &00000                         &2222  &  0.54   &0.61                   \\ \hline
July.2015          &00000                         &1111  &  0.55   &0.6                   \\ \hline
Aug.2015          &00000                         &2222  &  0.54   &0.6                   \\ \hline
Sep.2015          &00000                         &2222  &  0.55   &0.61                   \\ \hline
Oct.2015          &11111                         &0111  &  0.56   &0.62                   \\ \hline

\end{tabular}
\captionsetup{font=scriptsize,width=0.8\linewidth}
\caption*{The first column represents months with the lowest efficiency rates. Columns 2 and 3 are the most frequent blocks in 3- and 4-symbols discretization. Columns 4 and 5 are the probability of the success of the simple trading strategy for filtered and original price returns.}
\label{table:frequentblocksRSTI}
\end{table}

The simple trading strategy is an illustrative example of market inefficiency. In fact, such a strategy could result in no profit when used in practice because it does not take into account the costs of transaction and other trading frictions. Moreover, the filtering of daily seasonality pattern is made by using the whole period of analysis. That is, this method cannot be applied in real time. Finally, we consider blocks containing only observed returns, by neglecting the missing values from the analysis. Thus, the application of such a strategy in practice should be integrated with the case when a missing value follows a sequence of 3 symbols.
\clearpage
\subsection{Stock market clustering}
Most of the month-long time intervals are identified as inefficient. But is there some dependence between two stocks that are inefficient at the same time?
\subsubsection{Kullback–Leibler distance}
\begin{figure}[htb]
\centering
\includegraphics[width=11cm]{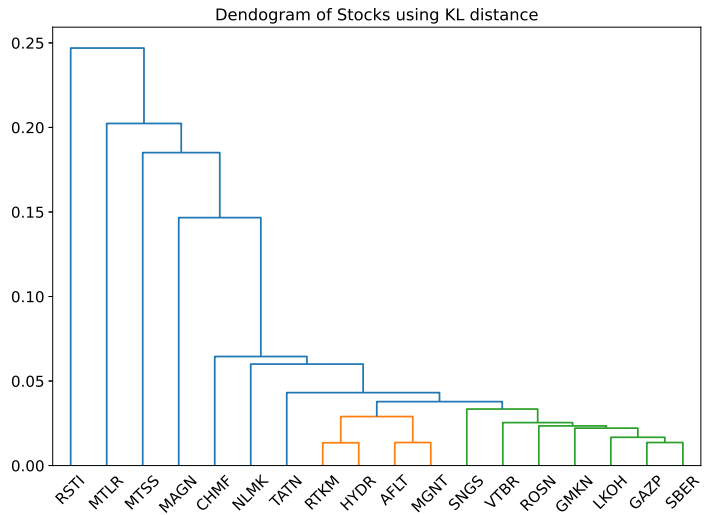}
\captionsetup{font=scriptsize}
\caption{Hierarchical clustering tree using KL distance. The threshold for clustering into groups is 0.035.}
\label{fig:Tree_average_1}
\end{figure}

We measure the similarity of discretized filtered returns by using the Kullback–Leibler (KL) distance (Eq.~\ref{KL}). We use $k$, the length of blocks, as the maximum value suitable for both sequences according to Eq.~\ref{value of k}. The 4-symbols discretization is used. The Kullback–Leibler divergence $DL(P|Q)$ is calculated using empirical frequencies. The entropy rates are calculated using Eq.~\ref{Eq: Grassberger}. Using the Kullback–Leibler distance for all pairs of stocks, we cluster them in three groups using hierarchical clustering with UPGMA algorithm \cite{Sokal}\footnote{This algorithm is implemented using the python function cluster.hierarchy.dendrogram with the argument distance=average.}. The result is in Fig.~\ref{fig:Tree_average_1}. Combining companies into one cluster means that their stocks have a common behavior that is not related to the value of volatility, the degree of price staleness and the structure of microstructure noise. 
It can be seen that banks and oil companies are clustered together (right). There is a group of four stocks RTKM, HYDR, AFLT, MGNT, that have nothing in common at first glance. The remaining group (left) mainly consists of metallurgy companies. However, there is no visible distinction between the stocks of banks and oil companies. According to the clustering tree, two telecommunications companies differ significantly, as well as electricity companies.

Finally, two stocks with the lowest efficiency rates, RSTI and MLTR, are the furthest (in the sense of KL distance) from any other stock. That is, there are no stocks that behave similarly to these two stocks.
\subsubsection{Entropy of co-movement}
\begin{figure}[htb]
\centering
\includegraphics[width=11cm]{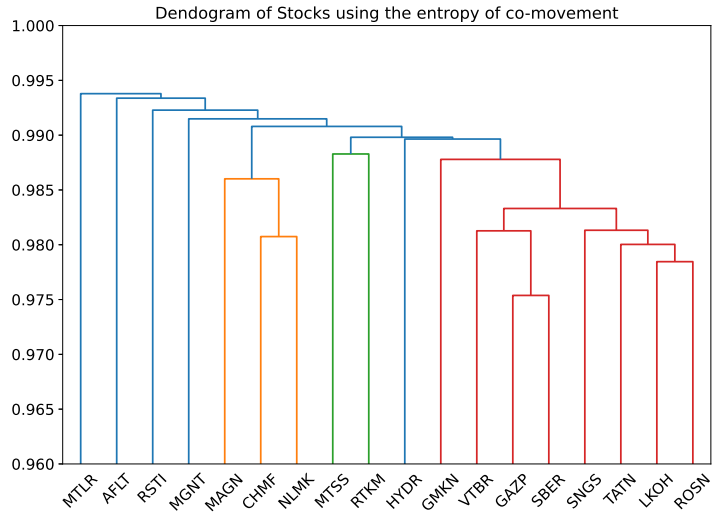}
\captionsetup{font=scriptsize}
\caption{Hierarchical clustering tree using the entropy of co-movement. The threshold for clustering into groups is 0.989.}
\label{fig:Tree_average_CoEntropy_1}
\end{figure}

Now, we consider another measure of difference between two stocks, the entropy of co-movement. We calculate the Shannon entropy of the discretization describing the movement of a pair of prices presented in Eq.~\ref{Eq:Comovement}. We consider only minutes that are in common for both stocks. For these minutes we consider values of residuals obtained after ARMA fitting. The result is in Fig.~\ref{fig:Tree_average_CoEntropy_1}.

Two companies related to telecommunications are a separate cluster. Three metallurgy companies MAGN, CHMF, NLMK also cluster together. Stocks relating to oil and bank companies form the other cluster. The same cluster, with the exception of the TATN (oil industry), was also formed in the previous section. The "closeness" of stocks GAZP and SBER is detected either in this and in the previous section. The three stocks on the left that join other stock clusters last are the stocks with the lowest efficiency rates.

Some clusters may form on the basis that companies belong to the same industry. The division of companies into industries is noticeable from the dendrogram in Figure~\ref{fig:Tree_average_CoEntropy_1}. However, this criterion does not explain all clusters. For instance, GMKN from metallurgy is in the cluster of oil companies and banks.
\section{Conclusions and discussion}
\label{Conclusions}
We have investigated the predictability of the Moscow Stock Exchange. We are interested in a measure of market inefficiency that is not related to known sources of regularity in financial time series. Usually, these sources are not filtered out and, accordingly, their impact is taken into account in the degree of price predictability (see e.g. \cite{Molgedey,Risso08,Mensi}).

We have focused on two sources of regularity, namely volatility clustering and price staleness \cite{Bandi}. Filtering of volatility clustering was made in \cite{Calcagnile} by estimating volatility using the exponentially weighted moving average. We have developed a modification of the volatility estimation by taking into consideration the effect of price staleness. Price staleness produces excess 0-returns that affect the estimation of volatility. Another approach of estimating volatility in the case of presence 0-returns was proposed in \cite{Sucarrat} where all 0-returns are reevaluated during an expectation-maximization algorithm. In our approach, we separate 0-returns that may have resulted from rounding and from price staleness. Thus, we also filter out apparent inefficiency due to price staleness. The advantage of our approach is simplicity: there is only one parameter in the method which can be optimized using historical data. Our approach combining the estimates of volatility and the degree of staleness can be used for real-time analysis since only past observations of time series are used.

We used the Shannon entropy as a measure of randomness to calculate a degree of inefficiency of the Moscow Stock Exchange. We used two types of the discretization of return time series to test efficiency more reliably for each month. The 4-symbols discretization helps to find more price movements that lead to market inefficiency than the 3-symbols discretization. There are $80\%$ of months over the period from 2012 to 2021 that are defined as inefficient. Even after filtering out all sources of apparent inefficiency, most of the months contain signals of market inefficiency.

We selected two stocks that exhibit the lowest efficiency rates. We have shown that the most inefficient months are grouped together. We have shown that, for such months, discretized price returns before and after filtering out apparent inefficiencies are predictable.

Finally, we used two methods to cluster stocks using filtered return time series. Inspired by \cite{Benedetto}, we computed the Kullback–Leibler distance between stocks and grouped stocks into three clusters. We also introduced the entropy of co-movement of two stocks. In this case, stock prices display common patterns that have an interpretation in terms of the sector the stocks belong to. We also noticed that the stocks of banks and oil companies were linked to each other. One possible improvement to stock clustering is to modify the entropy of co-movement such that it is possible to define a proper distance function. This is left for future research.

The proposed method for measuring market efficiency using the Shannon entropy can be applied in other markets of different countries. In this work, we use monthly time intervals for entropy calculation. Our future work will be related to the optimization of the length of return time series. One of the problems is to find a significant decrease in entropy without using Monte Carlo simulations. We also plan to switch to a higher frequency (less than one minute) to analyze the predictability of financial time series.
\begin{appendices}
\section{Data cleaning and whitening}
\label{Data cleaning and whitening}
\subsection{Outliers}
We use the method of an outlier detection introduced in \cite{Brownlees}. The algorithm finds price values that are too far from the mean in relation to the standard deviation. The algorithm deletes a price $P_i$ if
\begin{equation*}
|P_i-\bar{P}_i(k)|\ge c s_i(k)+\gamma,
\end{equation*}
where $\bar{P}_i(k)$ and $s_i(k)$ are respectively a $\delta$-trimmed sample mean and the standard deviation of the $k$ price records closest to time $i$. The $\delta\%$ of the lowest and the $\delta\%$ of the highest observations are discarded when the mean and the standard deviation are calculated from the sample. The parameters are $k = 20, \delta =5, c = 5, \gamma = 0.05$.
\label{Outliers}
\subsection{Stock Splits}
 We check the condition $|r| > 0.2$ in the return series to detect unadjusted splits\footnote{A split is a change in the number of company's shares and in the price of the single share, such that a market capitalization does not change.}. There are no unadjusted splits found.
\subsection{Intraday Volatility Pattern}
\label{Intraday Volatility Pattern}
The volatility of intraday returns has periodic behavior. The volatility is higher near the opening and the closing of the market. It shows an U-shaped profile every day. The intraday volatility pattern from the return series is filtered by using the following model. We define deseasonalized returns as
\begin{equation*}
\tilde{R}_{d,t}=\frac{R_{d,t}}{\xi_t},
\end{equation*}
where
\begin{equation*}
\xi_t=\frac{1}{N_{days}}\sum_{d'}\frac{|R_{d',t}|}{s_{d'}},
\end{equation*}
$R_{d,t}$ is the raw return of day $d$ and intraday time $t$, $s_d$ is the standard deviation of absolute returns of day $d$, $N_{days}$ is the number of days in the sample.
\subsection{Heteroskedasticity}
Different days have different levels of the deviation of the deseasonalized returns $\tilde{R}$. In order to remove this heteroskedasticity, we estimate the volatility $\bar{\sigma}_t$ in Appendix~\ref{appendix: Algorithm}. We define the standardized returns by
\begin{equation*}
    r_t=\frac{\tilde{R}_t}{\bar{\sigma}_t}.
\end{equation*}
\subsection{Price staleness}
If a transaction cost is high, the price is updated less frequently, even if trading volume is not zero. This effect is called price staleness and is discussed in Section~\ref{Estimation of price staleness}. We identify 0-returns appearing due to rounding (and not due to price staleness) using the Equation~\ref{staleness}. Other 0-returns are set as missing values as shown in Appendix~\ref{appendix: Algorithm}.
\subsection{Microstructure noise}
\label{Microstructure noise}
The last step in filtering apparent inefficiencies is filtering out microstructure noise. The microstructure effects are caused by transaction costs and price rounding. We consider the residuals of an ARMA(P,Q) model of the standardized returns after filtering out 0-returns. We apply the methodology introduced in \cite{Jones} to find the residuals of an ARMA(P,Q) model by using the Kalman filter. We select the values of $P$ and $Q$ that minimize the value of BIC \cite{Schwarz}, so that $P+Q<6$. The values of $P$ and $Q$ are chosen for each calendar year and are used for the next year. For the year 2012 we select $P=0$ and $Q=1$ corresponding to an MA(1) model.
\section{Algorithm}
\label{appendix: Algorithm}
The aim of the algorithm is to estimate volatility and filter out excess 0-returns due to price staleness. Some 0-returns appear due to price rounding. These 0-returns will be saved in the data. First, we set the number of 0-returns "to save" $N_{save}=0$ and the first value of a cumulative function $Z_1=0$. The cumulative function is updated $Z_t=Z_{t-1}+p_t$, if $r_{t-1}$ is not defined as missing due to staleness. Each time when $\lfloor Z(t) \rfloor - \lfloor Z(t-1)\rfloor=1$, $N_{save}$ is increased by $1$.

We notice that the first non-zero return after a row of 0-returns due to staleness is the sum of all missing returns generated by a hidden efficient price. This return is also set as missing. However, the value of return used for estimating volatility is calculated as its expected value: $\hat{r}_{n-1}=\frac{r_{n-1}}{\sqrt{N_0+1}}$, where $N_0$ is the amount of missing values strictly before the non-zero return $r_{n-1}$. The same is also referred to initially missing values, e.g., due to no-trading or errors in collecting the data.

Another assumption is that a 0-return appears due to staleness if the previous return had the 0-value and was defined to appear due to staleness. We include this rule, since we assume that it is more likely that two consecutive 0-returns appear due to high transaction costs than due to rounding (that is, simply speaking, two outcomes of generating Gaussian random variables are less than a tick size).

Generally, for the estimation of volatility at time $t$ we should consider three cases: $P_{t-1}$ was missing (or minute $t-1$ is non-trading), $r_{t-1}=0$, $r_{t-1}\ne 0$. Thus, the algorithm is the following. We give the algorithm for the case of $Sig_1$ which is used in the application for the real data. We remove all 0-returns that start the sequence.
\subsection{Pseudocode}
Step 0: $\bar{\sigma}_1=|r_1|/\mu_1$; $Z_1=0$, $N_{save}=0$; $N_0=0$.

For $t$ from $2$ to $N$, where $N$ is the length of time series:

Step 1:
\begin{itemize}
\item If $r_{t-1}$ is missing: $\bar{\sigma}_t=\bar{\sigma}_{t-1}$; Increase $N_0$ by the amount of consecutive missing prices 
\item Else if $r_{t-1}=0$: 
\begin{itemize}
\item If $N_{save}>0$ and  $N_0=0$: $N_{save}=N_{save}-1$, $\bar{\sigma}_t=Sig_1(\alpha, 0,\bar{\sigma}_{t-1})$
\item Else: $\bar{\sigma}_t=\bar{\sigma}_{t-1}$, $N_0=N_0+1$, $r_{t-1}=\text{missing}$
\end{itemize}
\item Else: $\bar{\sigma}_t=Sig_1(\alpha, \frac{r_{n-1}}{\sqrt{N_0+1}},\bar{\sigma}_{t-1})$, $N_0=0$
\end{itemize}
Step 2: 
\begin{itemize}
\item Calculate $p_t$ (Eq.~\ref{staleness})
\item If $r_{t-1}$ is not missing, $Z_t=Z_{t-1}+p_i$
\item If $\lfloor Z(t) \rfloor - \lfloor Z(t-1)\rfloor=1$, $N_{save}=N_{save}+1$
\end{itemize}

Finally, we check if the effect of staleness really exists in the price time series:
\begin{equation*}
\begin{split}
\hat{p}&=\frac{\sum_i p_i}{N}\\
\hat{q}&=1-\hat{p}\\
Var&=\hat{p}\hat{q}N
\end{split}
\end{equation*}
If $N_{real}\le \sum_i p_i+1.96\sqrt{Var}$, we leave time series without putting any missing values, where $N_{real}$ is the initial amount of 0-returns.
\section{A predictable time series with entropy at maximum}
\label{patterns in returns that are lost in discretization}
The goal of this section is to construct a price model where entropy is high because of discretization. This model shows that a high entropy value may be caused by discretization, but not because of the randomness of a return time series. 

There are equal probabilities of having symbols $0$, $1$ and $2$. $1$ corresponds to log-returns, $r$, equal to $-0.4$, $2$ corresponds to log-returns equal to $0.4$. The structure of symbol $0$ is more complicated. It covers three other symbols $3,4,5$. They correspond to log-returns $-0.3,0.1,0.2$, respectively. One of the symbols $3,4$ or $5$ appears with probabilities depending on the previous value of these symbols. The probabilities are presented in the Table~\ref{Table:Probabilities}. Having a symbol presented in a column, there are probabilities of getting a symbol presented in a row.
\renewcommand{\arraystretch}{3}
\begin{table}[ht]
\centering
\caption{Transition probabilities}
\begin{tabular}{|l|p{1cm}|p{1cm}|p{1cm}|}
\hline
  first symbol&$\cdot3$  &$\cdot4$  & $\cdot5$ \\ \hline
 $3\cdot$&$\dfrac{1}{6}$  &$\dfrac{1}{3}$  &$\dfrac{1}{2}$  \\ \hline
 $4\cdot$&$\dfrac{1}{2}$  &$\dfrac{1}{6}$  &$\dfrac{1}{3}$  \\ \hline
 $5\cdot$&$\dfrac{1}{3}$  &$\dfrac{1}{2}$  &$\dfrac{1}{6}$  \\ \hline
\end{tabular}
\captionsetup{font=scriptsize}
\caption*{Rows stand for the first symbol of a block, columns stand for the second symbol.}
\label{Table:Probabilities}
\end{table}

The model implies an average zero return. However, a trading strategy that increases a profit exists. After 3 a trader should buy, and after 4 and 5 the trader should sell. However, the entropy of 3-symbols series is at maximum, that should imply absence of profitable strategies.

Considering the same example with 4-symbols discretization we get that $Q_1=-0.4$, $Q_2=0.1$, and $Q_3=0.4$. Therefore, we have the following discretization of returns:
\begin{equation*}
s=
\begin{cases}
0, r=-0.4, \\
1, r=-0.3\text{ or }r=0.1,\\
2, r=0.2,\\
3, r=0.4.
\end{cases}
\end{equation*}

Thus, we can distinguish returns $r=0.2$ from others the using 4-symbols discretization. Table \ref{Table:Probabilities} gives the following probabilities for the blocks of two symbols from the 4-symbols discretization:
$p(11)=\frac{7}{162}$, $p(12)=p(21)=\frac{5}{162}$, $p(22)=\frac{1}{162}$. Noting that $p(0)=p(3)=\frac{1}{3}$, $p(1)=\frac{2}{9}$, and $p(2)=\frac{1}{9}$, we calculate that
$$H_1=-\frac{2}{3}\log\left(\frac{1}{3}\right)-\frac{2}{9}\log\left(\frac{2}{9}\right)-\frac{1}{9}\log\left(\frac{1}{9}\right)\approx0.946<1$$

and
\begin{equation*}
\begin{split}
H_2&=-\frac{1}{2}(\frac{7}{162}\log{\left(\frac{7}{162}\right)}+\frac{5}{81}\log{\left(\frac{5}{162}\right)}+\frac{1}{162}\log{\left(\frac{1}{162}\right)}+\\
&+\frac{4}{9}\log{\left(\frac{1}{9}\right)}+\frac{8}{27}\log{\left(\frac{2}{27}\right)}+\frac{4}{27}\log{\left(\frac{1}{27}\right)})\approx0.944<H_1
\end{split}
\end{equation*}
\end{appendices}
\clearpage
\bibliographystyle{unsrt}
\bibliography{file}
\end{document}